\documentclass[twocolumn,aps,prb,epsf,showpacs]{revtex4}
\usepackage{amsmath}
\usepackage{epsfig}
\usepackage{epsf}
\usepackage{array}

\begin{document}

\title{Interface ordering and phase competition 
in a model Mott-insulator--band-insulator heterostructure}
\author{Satoshi Okamoto}
\altaffiliation[Electronic address: ]{okapon@phys.columbia.edu}
\author{Andrew J. Millis}
\affiliation{Department of Physics, Columbia University, 538 West 120th Street, New York,
New York 10027, USA}
\date{\today }

\begin{abstract}
The phase diagram of model Mott-insulator--band-insulator heterostructures is
studied using the semiclassical approximation to the dynamical-mean-field method 
as a function of thickness, coupling constant, and charge confinement. 
An interface-stabilized ferromagnetic phase is found, allow the study of its competition and possible
coexistence with the antiferromagnetic order characteristic of the bulk Mott insulator. 
\end{abstract}

\maketitle


Fabrication and investigation of heterostructures involving correlated-electron 
materials~\cite{Imada98,Tokura00} are an important direction in material science. 
Understanding of the electronic properties near the interfaces and surfaces 
is not only of scientific interest 
but is also indispensable to realize electronic devices exploiting the unique properties of 
correlated-electron materials. 
A variety of heterostructures have been fabricated and studied including 
high-$T_c$ cuprates,\cite{Ahn99,Ahn02} 
Mott-insulator and band-insulator heterostructure,\cite{Ohtomo02} 
and superlattices of transition-metal oxides.\cite{Izumi01,Biswas00,Biswas01} 
Interestingly, the heterostructures comprising of a Mott-insulator and a band-insulator 
were reported to show metallic behavior.\cite{Ohtomo02} 

A fundamental question raised by those studies is 
``what electronic phases are realized at interfaces.'' 
For the vacuum-bulk interface (i.e., the surface), 
Potthoff and Nolting,\cite{Potthoff99} 
Schwieger {\it et al.},\cite{Schwieger03} and Liebsch~\cite{Liebsch03} have argued that 
the reduced coordination may enhance correlation effects. 
The enhanced correlations could presumably induce surface magnetic ordering, 
although this possibility was not discussed in Refs.~\onlinecite{Potthoff99,Schwieger03,Liebsch03}. 
Matzdorf {\it et al.} proposed that ferromagnetic ordering is stabilized at the surface 
of two-dimensional ruthenates by a lattice distortion,\cite{Matzdorf00} 
but this is not yet observed. 
Surface ferromagnetism had been also discussed in a mean field treatment of 
the Hubbard model by Potthoff and Nolting.\cite{Potthoff95} 
Similarly, the effect of bulk strain on the magnetic ordering in perovskite manganites 
was discussed by Fang {\it et al.}\cite{Fang00} 

All these studies dealt with systems in which charge densities remain unchanged from 
the bulk values, and physics arising from the modulation of charge density was not addressed. 
A crucial aspect of the recently fabricated heterostructures is charge inhomogeneity, 
caused by the spreading of electrons from one region to another. 
Strongly correlated materials typically possess interesting density-dependent phase diagrams;
raising the possibility of interesting phase behavior at interfaces. 
In this paper we use the semiclassical approximation (SCA)\cite{Okamoto05} 
to the dynamical-mean-field method\cite{Georges96} 
to explore the phase behavior of a model Mott-insulator--band-insulator heterostructure. 
The SCA is computationally inexpensive and a good representation of phase diagrams
and transition temperatures in several models, allowing us to investigate the phase behavior 
for a wide range of parameters, and in particular to access the $T>0$ regime which Hartree-Fock and 
related approximations fail to represent adequately.\cite{Okamoto04a} 
We observe antiferromagnetic ordering in regions with charge density $\sim 1$ characteristic of 
a bulk Mott insulator, while 
ferromagnetic ordering is found to be a surface effect supported by an intermediate charge density
and a strong coupling. 
However, we have found that, despite the successes noted in previous work, 
the SCA overestimates ferromagnetism on the lattice we study here. 
Therefore, our results should be regarded as qualitative explanations of the type of phase behavior
which may occur rather than as quantitative statements about the Hubbard-model phase diagram. 

We study the model heterostructure introduced in Ref.~\onlinecite{Okamoto04b}.  
The Hamiltonian is a simplified representation of conduction bands of the systems studied 
in Ref.~\onlinecite{Ohtomo02} with the orbital degeneracy neglected. 
We consider [001] heterostructures formed by varying the $A$-site of a $AB$O$_3$ perovskite lattice. 
The electrons of interest reside on the $B$-site ions, which form a simple cubic lattice with sites 
labeled by $i$ as $\vec{r}_{i}=(x_{i},y_{i},z_{i})=a(n_{i},m_{i},l_{i})$ 
with the lattice constant $a$ set to unity. 
We assume each $B$-site has a single orbital; 
electrons hop between nearest neighbor sites with the transfer $t$. 
The electrons interact via a on-site interaction $U$ and a long-ranged Coulomb repulsion. 
The heterostructure is defined by $n$ planes of charge +1 counterions placed on 
the $A'$ sublattice of $A$-site ions at positions $\vec{r}_{j}^{A'}=a(n_{j}+1/2,m_{j}+1/2,l_{j}+1/2)$, 
with $-\infty<n_{j},m_{j}<\infty $ and $l_{j} = 1, \ldots , n$. 
(Coordinate $z$ will be shifted such that the center of the heterostructure, 
$A'$ sublattice, comes to $z=0$.)
Charge neutrality requires that the areal density of electrons is $n$. 
The resulting Hamiltonian is 
$H = H_{band}+H_{int}+H_{Coul}$ with 
\begin{eqnarray}
H_{band} &=&-t\sum_{\langle ij\rangle, \sigma}(d_{i\sigma }^{\dag }d_{j\sigma
}+H.c.),  \label{Hband} \\
H_{int} &=& U\sum_{i}n_{i\uparrow }n_{i\downarrow } +\frac{1}{2}\sum_{
{i\neq j} \atop {\sigma, \sigma^{\prime}}} 
\frac{e^{2}n_{i \sigma}n_{j
\sigma^{\prime}}}{\varepsilon |\vec{r}_{i}-\vec{r}_{j}|},  \label{Hint} \\
H_{Coul} &=& -\sum_{i,j, \sigma}\frac{e^{2}n_{i \sigma}} 
{\varepsilon |\vec{r}_{i}-\vec{r}_{j}^{A'}|}.  \label{Hcoul}
\end{eqnarray}
Note that $U \ne 0$ on all sites.  
A dimensionless measure of the strength of the long-ranged Coulomb interaction is 
$E_c=e^2/(\varepsilon a t)$ with the dielectric constant $\varepsilon$. 
In most of our analysis, we choose $E_c=0.8$. 
This corresponds to $t \sim 0.3$~eV, $a \sim 4$~\AA, and $\varepsilon=15$, 
which describe the system studied in Ref.~\onlinecite{Ohtomo02}. 
The charge profile is found not to depend in an important way on 
$\varepsilon$, but the stability of magnetic orderings does 
because this is sensitive to the details of the charge density distribution 
as discussed later. 

The basic object of our study is the electron Green's function. 
In general, this is given by 
\begin{equation}
G_\sigma(\vec r, \vec {r'} ;\omega )=[\omega +\mu
-H_{band}-H_{Coul}-\Sigma_\sigma (\vec r, \vec {r'} ;\omega )]^{-1}, 
\label{eq:Greenlatt}
\end{equation}%
with the chemical potential $\mu$ and the electron self-energy $\Sigma$. 
We consider [001] heterostructures with either in-plane translational invariance or 
$N_s$-sublattice antiferromagnetism. 
The Green's function and self-energy are therefore functions of the variables 
$(z, \eta, z', \eta', \vec{k}_\parallel)$ where $\eta$ and $\eta' (=1,\ldots,N_s)$ label the sublattice 
in layers $z$ and $z'$, respectively, and $\vec{k}_\parallel$ is a momentum in the (reduced) 
Brillouin zone. 
As in Ref.~\onlinecite{Okamoto04b}, 
we approximate the self-energy as the sum of a static Hartree term $\Sigma_\sigma^H$ 
arising from the long-ranged part of the Coulomb interaction  
and a dynamical part $\Sigma_\sigma^D (\omega)$ arising from local fluctuations.
Generalizing the inhomogeneous case of the dynamical-mean-field theory (DMFT),\cite{Schwieger03} 
we assume that the self-energy is only dependent on layer $z$ and sublattice $\eta$. 
Thus, the dynamical part of the self-energy is written as 
\begin{equation}
\Sigma_\sigma^D\Rightarrow \Sigma_\sigma^D (z, \eta; \omega ).  \label{sigd}
\end{equation}
The $z \eta$-dependent self-energy is determined from the solution of 
a quantum impurity model~\cite{Georges96} 
with the mean-field function fixed by the self-consistency condition 
\begin{equation}
G_\sigma^{imp}(z, \eta; \omega)= N_s \int \!\! \frac{d^{2}k_{\parallel}}{(2\pi )^{2}} \,\,
G_\sigma (z, \eta, z, \eta, \vec{k}_{\parallel };\omega ).  \label{sce}
\end{equation}

We shall use the DMFT, which is a mean-field approximation, to calculate magnetic phase diagrams 
of effectively two dimensional systems, so some mention of fluctuation effects is needed.
In general, the calculated transition temperatures are to be understood as crossover scales 
below which the magnetic correlation length $\xi$ grows rapidly $\xi \sim \exp(2 \pi \rho_s/T)$ with 
spin stiffness $\rho_s$~\cite{Sachdev99} discussed in more detail below.
True long-ranged order may be induced at $T>0$ by an Ising anisotropy or by coupling in a system 
made of repeated heterostructure units. 
In any event, the rapid growth of the correlations for $T<\rho_s$ means that the properties are effectively 
those of an ordered state. 
We will find that the low-$T$ $\rho_s$ is sufficiently large relative to the calculated transition temperature
that fluctuation effects are not crucial. 

In general for the heterostructure with $L$ layers with $N_s$ sublattices, 
one must solve $L \times Ns$ independent impurity models. 
Due to the self-consistency condition [cf. Eq.~(\ref{sce})] 
and to compute the charge density $n_\sigma (z, \eta)=- \int \frac{d\omega}{\pi} 
f(\omega) \mathrm{Im} G_\sigma^{imp}(z, \eta; \omega )$ with $f$ the Fermi distribution function, 
it is required to invert the $(L \times Ns)^2$ Green's function matrix at each momenta and frequency. 
This time consuming numerics restricts the size of the unit cell. 
In this study, we consider the commensurate magnetic states with up to two sublattices, 
$N_s =1$ and 2, on each layer and with the charge density independent of the sublattices, i.e., 
paramagnetic (PM), ferromagnetic (FM) states, and (layer-) antiferromagnetic (AF) state 
where antiferromagnetic (FM) planes with moment alternating from plane to plane. 
Note that the AF state extrapolates to the bulk AF state with the magnetic vector 
$\vec q =(\pi,\pi,\pi)$ at $n \rightarrow \infty$. 
By symmetry, the number of quantum impurity models one must solve is reduced to $L$. 
However, solution of the impurity models is a time consuming task, 
and an inexpensive solver is required. 
In Ref.~\onlinecite{Okamoto04b}, to study the evolution of 
the low-energy quasiparticle band and high-energy Hubbard bands 
as a function of position, we applied two-site DMFT~\cite{Potthoff01}
which is a simplified version of exact-diagonalization method. 
At $T=0$, this method is known to give reasonable result for Mott metal-insulator and 
magnetic transitions. 
However, small number of bath orbitals is known to be insufficient 
to describe the thermodynamics correctly.\cite{Okamoto05} 

For the investigations presented here, we use the semiclassical approximation, 
which is computationally inexpensive and has been found to be reasonably accurate 
for phase boundaries and excitation spectra of several models, including the 
half-filled Hubbard model and, at all fillings, for 
the $d=3$ and $d=\infty$ face-centered-cubic (FCC) lattices.\cite{Okamoto05}
We note however that, perhaps because it does not properly include quasiparticle coherence, 
the method overemphasizes ferromagnetism at intermediate density, giving for the FCC lattices 
transition temperatures $\sim 50\,\%$ higher than those found by quantum Monte-Carlo,\cite{Okamoto05} 
and for cubic lattices, finding ferromagnetism at $n_{tot}=0.5$ and moderately large $U$ 
(of order the critical value for the Mott transition) when other methods\cite{Zitzler02} 
suggest ferromagnetism is confined to very large $U$ and $n_{tot}$ near 1. 
Thus, the application of the SCA to Hubbard heterostructure allows a convenient exploration of 
the interplay between different phases, but probably does not provide a quantitatively 
reliable picture of the phase diagram. 

\begin{figure}[tbp]
\includegraphics[width=0.9\columnwidth,clip]{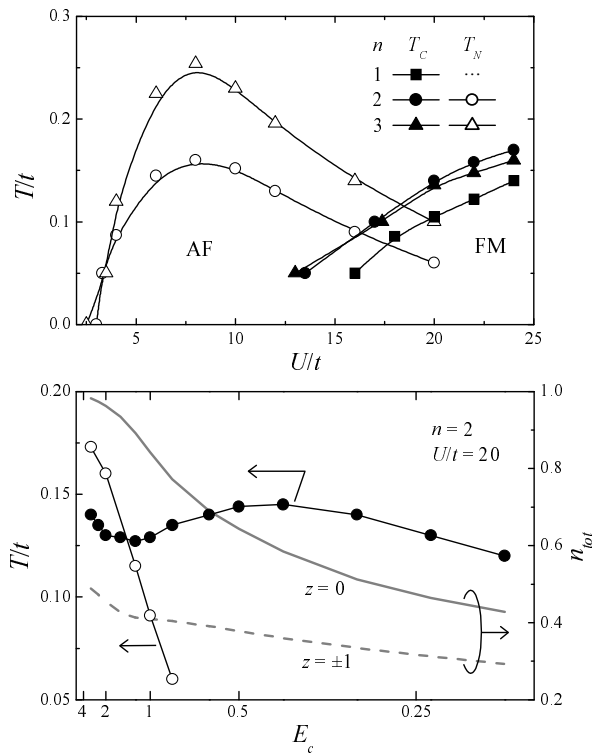}
\caption{Upper panel: 
Magnetic transition temperatures of heterostructures with various thicknesses $n$ indicated 
as functions of interaction strength. $E_c=0.8$. 
Filled symbols: Ferromagnetic Curie temperature $T_C$, open symbols: 
Antiferromagnetic N{\' e}el temperature $T_N$. 
Note that, where both phases are locally stable, 
the phase with higher $T_C (T_N)$ has the lower free energy. 
Lower panel: Magnetic transition temperatures and charge density $n_{tot}$ 
at $z=0$ (light solid line) and $\pm 1$ (light broken line) at $T/t=0.1$ 
as functions of the parameter $E_c$ 
for 2-layer heterostructure with $U/t=20$. Counterions are placed at $z= \pm 0.5$. 
}
\label{fig:PDfiniteT}
\end{figure}

First, we investigate the magnetic behavior at finite temperature. 
The upper panel of Fig.~\ref{fig:PDfiniteT} shows our calculated phase diagram in 
the interaction-temperature plane for heterostructures with various thicknesses for charge binding
parameter $E_c=0.8$. 
The one-layer heterostructure is PM at weak to moderate interactions, and FM at strong interactions. 
The two- and three-layer heterostructures are AF at weak to intermediate interaction, 
and become FM at stronger interactions with almost the same $T_C$ for $n=2$ and 3. 
Antiferromagnetic N{\'e}el temperature $T_N$ is found to be strongly dependent on the layer thickness; 
it increases with the increase of layer thickness.  
These $T_N$'s are substantioally reduced from the the bulk values; 
maximum value $T_N^{max}/t \sim 0.47$ at $U/t \sim 10$, whereas Curie temperature 
$T_C$'s are almost the bulk 2$d$ values at $n_{tot} \sim 0.5$. 
The strong dependence of $T_N$ on thickness may be understood from the bulk phase diagram;
antiferromagnetism is stabilized only very near to half-filling, and in the thinner heterostructures 
the charge-spreading effect reduces the density too much. 
This physics is seem from a different point of view in the lower panel of Fig.~\ref{fig:PDfiniteT}. 
Hartree-Fock studies of this and related models find a layer-AF phase. 
This phase is not found in our DMFT analysis. 

The lower panel of Fig.~\ref{fig:PDfiniteT} presents a detailed study of the $n=2$ heterostructure 
showing how changes in the charge confinement parameter $E_c$ affect the physics. 
The filled and open points (left-hand axis) show the variation of the Curie and N{\'e}el temperatures, 
respectively. 
The light solid and light broken lines (right-hand axis) show the variation of charge density 
on the central and next to central layers, respectively. 
It is seen that the AF ordering is rapidly destabilized 
with the decrease of $E_c$ (and concomitant decrease of charge density),  
and $T_N$ is seen to be correlated to that of the charge density at $z=0$. 
On the contrary, $T_C$ has a weak variation with $E_c$. 
This indicates that the FM ordering is favored by the intermediate charge density 
as discussed in the bulk single-band Hubbard model;\cite{Denteneer95} 
at large $E_c$, the magnetization is large on the outer layer and small in the inner layer, 
at small $E_c$, the situation is reversed. 

\begin{figure}[tbp]
\includegraphics[width=0.9\columnwidth,clip]{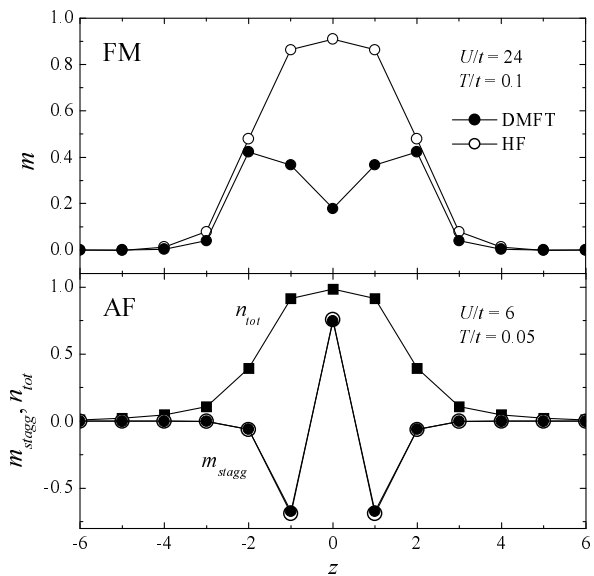}
\caption{Magnetization density of 4-layer heterostructure. 
Counterions are placed at $z= \pm 0.5, \pm 1.5$. $E_c=0.8$. 
Upper panel: Magnetization $m$ in a FM state for $U/t=24$ and $T/t=0.1$. 
Lower panel: In-plane staggered magnetization $m_{stagg}$ in an AF state for 
$U/t=6$ and $T/t=0.05$. 
Filled (open) circles are the results by DMFT (HF). 
For comparison, charge density $n_{tot}$ computed by DMFT is also shown in 
the lower panel (filled squares). 
Note that the staggered magnetization in the outer layers ($|z| \ge 2$), 
and the outermost layers ($|z|=1$) have the same sign. 
}
\label{fig:NandM}
\end{figure}

These arguments are confirmed by our calculations of the spatial variation of the magnetization density,  
reported in Fig.~\ref{fig:NandM}, which shows 
numerical results for a 4-layer heterostructure with counterions at $z=\pm 0.5$ and $\pm1.5$, and $E_c=0.8$. 
The upper panel of Fig.~\ref{fig:NandM} shows the magnetization in the FM state. 
In DMFT (filled circles), only the layers near the interfaces ($|z| \sim 1$--2) have large polarization 
and inner layers in the heterostructure have small moments. 
This explains the weak $n$-dependence of $T_C$ of thick heterostructures 
(see the upper panel of Fig.~\ref{fig:PDfiniteT}). 
In HF (open circles), all layers in the heterostructure are highly polarized. 
In contrast in an AF state, result of the in-plane staggered magnetization by DMFT and HF agree well 
as shown in the lower panel of Fig.~\ref{fig:NandM}. 
For comparison, the total charge density is also plotted (filled squares). 
The in-plane staggered magnetization is large only at inner layers where the charge density is close to 1. 
Note that the staggered magnetization in the outer layers ($|z| \ge 2$) has the same sign 
as in the outermost layers ($|z|=1$) indicating that the outer layers are not intrinsically magnetic. 

We now return to the issue of fluctuation effects. 
We study a model which is two dimensional and order parameters with a continuous spin rotation symmetry;
thus, at $T>0$, rather the calculated $T_C$ means a crossover to a low-$T$ ``almost ordered'' phase 
characterized by a exponentially growing correlation length $\xi \sim \exp(2 \pi \rho_s/T)$\cite{Sachdev99} 
where the key stiffness $\rho_s$ is given by the second derivative of the free energy with respect to 
the order parameter orientation. 
We have computed $\rho_s$ for a single ferromagnetic plane with charge density 0.5 finding
$\rho_s \sim t/8$ 
We observe that in all cases several planes exhibit the relevant order, so that the total stiffness
is larger by a factor 2--3. 
Thus we see that the stiffness is large enough relative to the DMFT $T_{C,N}$'s that 
the purely two dimensional fluctuation effects are of minor importance, leading to $\xi > \exp (4$--$5)$ 
for $T<T_{C,N}/2$. 

\begin{figure}[tbp]
\includegraphics[width=0.9\columnwidth,clip]{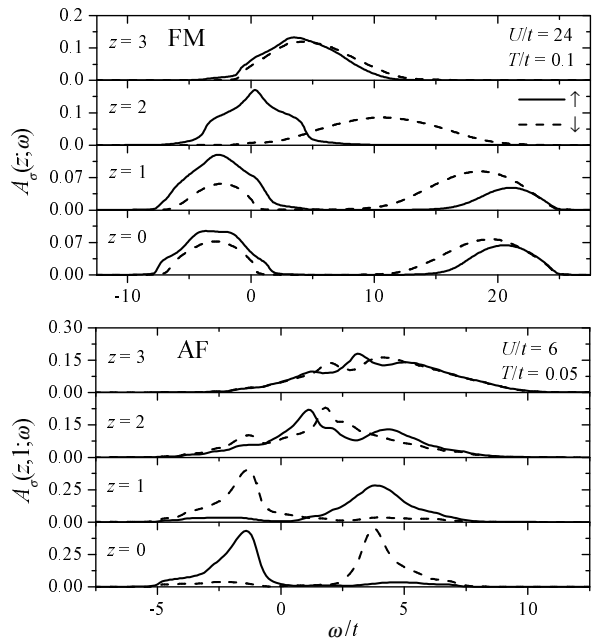}
\caption{Layer- and sublattice-resolved spectral functions as functions of real frequency $\omega$
for 4-layer heterostructure. $E_c=0.8$. 
Upper panel: Ferromagnetic state at $U/t=24$ and $T/t=0.1$. 
Lower panel: Antiferromagnetic state at $U/t=6$ and $T/t=0.05$. Sublattice $\eta =1$. 
Solid (broken) lines are for up (down) spin electrons. 
For sublattice $\eta =2$ in AF state, up and down electrons are interchanged. 
}
\label{fig:spectra}
\end{figure}

Spin distributions presented in Fig.~\ref{fig:NandM} can be understood from the 
single-particle spectral functions. 
In Fig.~\ref{fig:spectra} are presented the DMFT results for 
the layer- and sublattice-resolved spectral functions 
$A_\sigma(z, \eta; \omega)= - \frac{1}{\pi} \mathrm{Im} G_\sigma^{imp}(z,\eta; \omega +i0^{+})$ 
for the FM (upper panel) and the AF (lower panel) states 
of 4-layer heterostructure with the same parameters as in Fig.~\ref{fig:NandM}. 
These quantities can in principle be measured by 
spin-dependent photoemission or scanning tunneling microscopy. 
As noticed in Ref.~\onlinecite{Okamoto04b}, 
spectral function outside of the heterostructure ($|z| \gg 2$) is essentially identical 
to that of the free tight-binding model $H_{band}$, and electron density is negligibly small. 
With approaching the interfaces ($|z|=2$), 
the spectral function shifts downwards and begins to broaden. 
In the FM case, magnetic ordering is possible only near the interface ($|z| \sim 1$--2) 
carrying the intermediate charge density. 
Inside the heterostructure ($|z| \ll 2$), clear Hubbard gap exists due to the large $U$ 
and uniform polarization is hard to achieve. 
On the contrary, high charge density is necessary to keep the staggered magnetization 
in the AF case as seen as a difference between up and down spectra in the lower panel of 
Fig.~\ref{fig:spectra}. 

So far, we have discussed the competition between ferromagnetic and antiferromagnetic phases 
and spatial distributions of charge and magnetic densities establishing 
that ferromagnetism is a surface effect. 
Finally, we discuss the interplay between phases. 
We note that the charge density or magnetization density typically varies over $\sim3$ unit cell range, 
so for thin heterostructures and moderate charge confinement energies, the entire heterostructure exhibits 
a single phase, controlled by the instability exhibiting the highest transition temperature 
(cf. Fig.~\ref{fig:PDfiniteT}). 
However, for thicker heterostructure or stronger confinement we believe that one can observe an ordered state
involving an antiferromagnetic center with a ferrimagnetic ``skin.''
A hint of this behavior can be observed in Fig.~\ref{fig:NandM}.
Consider the central layer; this has an occupancy $n_{tot} \approx 0.9$, so from Fig.~\ref{fig:PDfiniteT} 
and rescaling account for the different $U$, we would expect $T_N \sim 0.1t$. 
Now this layer is not isolated; hoppings to the layers of $z=\pm1$ leads to a effective polarizing field 
of the order of $2 (1-n_{tot}) t \, m/n_{tot}$ where the first factor is the number of layers, 
the second is the hopping amplitude renormalized by strong correlations, and the third factor 
is the relative spin polarization; putting these factors together gives a polarizing field of about $0.15t$, 
approximately equal to the AF coupling. 
Thus for the central layer of this heterostructure ferromagnetic and antiferromagnetic tendencies are very 
closely balanced, but for thicker systems (not at present computationally accessible) or perhaps for stronger 
charge confinement, an antiferromagnetic center will occur. 

To summarize, we have presented a semiclassical DMFT study of 
magnetic phase behavior of a model Mott-insulator--band-insulator heterostructure  
in which the behavior is controlled by the spreading of the electronic charge out of the confinement region. 
Magnetic phase diagram is investigated as a function of layer thickness, temperature, and interaction strength. 
Ferromagnetic ordering is found to be a surface effect 
stabilized at an interface region with moderate charge density, 
while antiferromanetic ordering is found at a region with high density $\sim 1$ 
characteristic of the bulk Mott insulator, 
and N{\' e}el temperature is sensitive to the layer thickness and charge confinement energy. 
These magnetic orderings may coexist in very thick heterostructure 
exhibiting a ferromagnetic ``skin'' and an antiferromagnetic ``core.'' 

We acknowledge fruitful discussions with G. Kotliar, P. Sun, J. Chakhalian, 
D. Vollhardt, and Th. Pruschke. 
This research was supported by JSPS (S.O.) and the DOE under Grant No. ER 46169 (A.J.M.).


\begin{thebibliography}{99}

\bibitem{Imada98} M.~Imada, A. Fujimori, and Y. Tokura, Rev. Mod. Phys. 
\textbf{70}, 1039 (1998).

\bibitem{Tokura00} Y. Tokura and N. Nagaosa, Science \textbf{288}, 462
(2000).

\bibitem{Ahn99}C. H. Ahn, S. Gariglio, P. Paruch, T. Tybell, L. Antognazza, and 
J.-M. Triscone, Science, \textbf{284}, 1152 (1999).

\bibitem{Ahn02}S. Gariglio, C. H. Ahn, D. Matthey, and J.-M. Triscone, 
Phys. Rev. Lett. \textbf{88}, 067002 (2002).

\bibitem{Ohtomo02} A. Ohtomo, D. A. Muller, J. L. Grazul, and H. Y. Hwang,
Nature \textbf{419}, 378 (2002).

\bibitem{Izumi01} M. Izumi, Y. Ogimoto, Y. Konishi, T. Manako, M. Kawasaki,
and Y. Tokura, Mat. Sci. Eng. B \textbf{84}, 53 (2001) and references
therein.

\bibitem{Biswas00}A. Biswas, M. Rajeswari, R. C. Srivastava, Y. H. Li, 
T. Venkatesan, R. L. Greene, and A. J. Millis, Phys. Rev. B \textbf{61}, 9665 (2000).

\bibitem{Biswas01}A. Biswas, M. Rajeswari, R. C. Srivastava, T. Venkatesan,
R. L. Greene, Q. Lu, A. L. deLozanne, and A. J. Millis, 
Phys. Rev. B \textbf{63}, 184424 (2001).



\bibitem{Potthoff99}M. Potthoff and W. Nolting, 
Phys. Rev. B \textbf{60}, 7834 (1999).

\bibitem{Schwieger03} S. Schwieger, M. Potthoff, and W. Nolting, 
Phys. Rev. B \textbf{67}, 165408 (2003).

\bibitem{Liebsch03}A. Liebsch, Phys. Rev. Lett. {\bf 90}, 096401 (2003). 

\bibitem{Matzdorf00}R. Matzdorf, Z. Fang, Ismail, J. Zhang, T. Kimura, 
Y. Tokura, K. Terakura, and E. W. Plummer, Science \textbf{289}, 746 (2000).

\bibitem{Potthoff95}M. Potthoff and W. Nolting, Phys. Rev. B {\bf 52}, 15341 (1995).

\bibitem{Fang00}Z. Fang, I. V. Solovyev, and K. Terakura, 
Phys. Rev. Lett. \textbf{84}, 3169 (2000).

\bibitem{Okamoto05}S. Okamoto, A. Fuhrmann, A. Comanac, and A. J. Millis, 
Phys. Rev. B {\bf 71}, 235113 (2005). 

\bibitem{Georges96}A. Georges, B. G. Kotliar, W. Krauth, and M. J. Rozenberg, 
Rev. Mod. Phys. \textbf{68}, 13 (1996).

\bibitem{Okamoto04a}S. Okamoto and A. J. Millis, Nature (London) \textbf{428}, 630; 
Phys. Rev. B {\bf 70}, 075101 (2004).

\bibitem{Okamoto04b}S. Okamoto and A. J. Millis, 
Phys. Rev. B {\bf 70}, 241104(R) (2004). 

\bibitem{Sachdev99}S. Sachdev, {\it Quantum Phase Transitions} (Cambridge University Press, Cambridge, 1999). 

\bibitem{Potthoff01}M. Potthoff, Phys. Rev. B \textbf{64}, 165114 (2001).

\bibitem{Zitzler02}R. Zitzler, T. Pruschke, and R. Bulla, Eur. Phys. J. B {\bf 27}, 473 (2002). 

\bibitem{Denteneer95}P. J. H. Denteneer and M. Blaauboer, 
J. Phys.: Condens. Matter {\bf 7}, 151 (1995). 


\end{thebibliography}
\end{document}